\documentclass[twocolumn,english,showpacs,preprintnumbers,amsmath,amssymb,floatfix,nofootinbib]{revtex4-1}

\usepackage[T1]{fontenc}
\usepackage[latin9]{inputenc}
\usepackage{color}
\usepackage{array}
\usepackage{amstext}
\usepackage{graphicx}
\usepackage{esint}
\usepackage{rotating}
\usepackage[font=small,labelfont=bf]{caption}
\usepackage{xcolor}
\usepackage[colorlinks = true,
linkcolor = magenta,
urlcolor  = blue,
citecolor = red,
anchorcolor = blue]{hyperref}

\usepackage{float}

\newcommand{\A}{{\mathcal{A}}}
\newcommand{\tA}{{\widetilde {\mathcal{A}}}}

\newcommand{\tk}{{\widetilde k}}

\def\as{\relax\ifmmode \alpha_s\else{$ \alpha_s${ }}\fi}  
\def\abar{\relax\ifmmode{\bar{a}}\else{$\bar{a}${ }}\fi}  
\def\be{\begin{equation}}
\def\ee{\end{equation}}
\def\ba{\begin{eqnarray}}
\def\ea{\end{eqnarray}}

\makeatletter

\providecommand{\tabularnewline}{\\}

\@ifundefined{textcolor}{}
{%
 \definecolor{BLACK}{gray}{0}
 \definecolor{WHITE}{gray}{1}
 \definecolor{RED}{rgb}{1,0,0}
 \definecolor{GREEN}{rgb}{0,1,0}
 \definecolor{BLUE}{rgb}{0,0,1}
 \definecolor{CYAN}{cmyk}{1,0,0,0}
 \definecolor{MAGENTA}{cmyk}{0,1,0,0}
 \definecolor{YELLOW}{cmyk}{0,0,1,0}
 }

\@ifundefined{definecolor}
 {\usepackage{color}}{}
\@ifundefined{definecolor}
 {\usepackage{color}}{}
\makeatother
\usepackage{babel}

\begin{document}


\title {\textcolor[rgb]{0.00,0.00,0.00}{Nonsinglet polarized nucleon structure function in infrared-safe QCD}}

\author { Leila Ghasemzadeh$^{1}$}
\thanks{Electronic address}
\email{Leilaghasemzadeh@stu.yazd.ac.ir}

\author{ Abolfazl Mirjalili$^{1}$}
\thanks{Corresponding author}
\email{A.Mirjalili@yazd.ac.ir}

\author {S. Atashbar Tehrani$^{2}$}
\thanks{Electronic address}
\email{Atashbar@ipm.ir}

\affiliation {
$^{(1)}$Physics Department, Yazd University, P.O.Box 89195-741, Yazd, Iran       \\
$^{(2)}$School of Particles and Accelerators, Institute for Research in Fundamental Sciences (IPM), P.O.Box
19395-5531, Tehran, Iran}

\date{\today}

%
\begin{abstract}\label{abstract}
The  polarized nucleon structure function in the nonsinglet case  is investigated here by a new insight rather than  conventional perturbative QCD (pQCD).
For this purpose we note  that the solution of the evolution equations in moment space involves noninteger powers of the coupling constant. Therefore it is possible to employ a new approach which is called fractional analytical perturbation theory Consequently, it is possible to remove the Landau singularities of the renormalized coupling, i.e., at the scales $Q \sim \Lambda$, using this approach. This provides
an opportunity to continue the desired calculations toward small values of energy scales even less than the $\Lambda$ scale. To modify the analytical
perturbation theory, a newer approach is introduced, called 2$\delta$anQCD, in which the spectral function of the holomorphic coupling is
parameterized in the low-energy region by two delta  functions. This model gives us more reliable results for the considered QCD observables,
even in the deep infrared region. We calculate the nonsinglet part of the polarized nucleon structure function, using the 2$\delta$anQCD
model, and compare it with the result from the underlying pQCD where both are in a new defined scheme, called the Lambert scheme. For this purpose we employ the anQCD package in the \textit{Mathematica} environment to
establish the analytic (holomorphic) coupling constant. The results at various energy scales are also compared with the available experimental data, and it
turns out that there is a good consistency between them. The results show that the obtained nucleon structure function at small energy scales has smoother
behavior when using the 2$\delta$anQCD model than the underlying pQCD.  In fact  the  coupling constant  in  analytic QCD  behaves moderately and it  makes the result approach  the available data in a better way.
We also consider the $Q^2$ dependence of the Bjorken sum rule the (BSR), using  the 2$\delta$anQCD model. There is a good agreement between the
available experimental data for BSR and the  results from the utilized model, especially at low-energy scales.
\end{abstract}

\maketitle

%
\section{Introduction}\label{Introduction}
{{} QCD analysis of deep-inelastic scattering (DIS) data provides us new insight into the hadron physics. The reliability of theoretical results can be increased by testing the analysis, for instance, of hard lepton scattering off the hadrons. A reliable description of DIS at large momentum transfer $Q^2 \gg 1$ GeV$^2$ can be obtained, considering the twist expansion and factorization theorem. Following that one can use the DIS processes to better understand  the nucleon structure function. In this regard the polarized and unpolarized structure functions can be first modeled, using different parametrization methods. Then they can be evolved to high-energy scales, employing the QCD evolution equations. During the evolution process  the  numerical values of structure functions are affected by the energy scale, at which the renormalized coupling constant as the perturbation expansion parameter is evaluated. On the other hand the analysis of the structure functions should lead to consistent results with the related experimental data. To achieve this aim it is needed to model the parton densities, as ingredients of nucleon structure, at an initial energy scale. A further step is the evolution of these densities to high-energy scales. Therefore, the role of the
renormalized coupling constant $\alpha_s(Q^2)$ is important and should be considered in detail.

At low-$Q^2$  energy scales, below 1 or 2 ${\rm GeV}^2$, the qualification of DIS as a hard process is not appropriate. Its analysis is faced with two essential difficulties in the low-$Q^2$ region. One of them is related to the higher-twist corrections which  are dominantly affecting  the contribution of the
leading twist at low values of the energy scale. Therefore, the twist corrections at higher orders in $\alpha_s$ but at a low-energy scale causes the first difficulty.
In addition, since $\alpha_s(Q^2)$ as the QCD running coupling  grows rapidly at the energy scales near the Landau singularities, the result of  pQCD would not be reliable and this represents the second difficulty.

A new approach that is called the analytic perturbation theory (APT) can be applied to resolve these problems. This approach has been developed by Shirkov, Solovtsov {\it et al.\/} \cite{ShS,MS,MSS,MSa,Sh1,Sh2}. There, the running QCD coupling of conventional pQCD is transformed into an analytic (holomorphic) function of $Q^2$ which is accordingly called the APT coupling constant. In this regard, one needs a dispersion relation, involving a spectral or
discontinuity function which is the imaginary part of the coupling along the timelike semiaxis ($Q^2<0$) in the complex $Q^2$ plane.  In the framework of APT,
the images (analogs) which correspond to integer powers  of the original perturbative QCD (pQCD) coupling constant can  also be constructed. Later on, an extension to the analogs of noninteger powers was performed, which is called fractional APT (FAPT).
Some newer constructions of the coupling constant in analytic QCD (anQCD) are  based on the modifications of the spectral  function
at low-energy scales. In the case of this modification, the spectral function is constructed in a specific manner, {by parameterizing its behavior in the low-energy region by two positive delta functions \cite{CCEMAyala,Ayala:2014pha}.} The presented results of this paper are based on employing this model which is called  $2\delta$anQCD.}

We remark that anQCD is not the only approach to obtain a proper behavior of the coupling constant at low energy, i.e., at infrared (IR) scales. To achieve  an IR-safe QCD description one can also refer specifically to the AdS/CFT coupling of Brodsky, de Teramond, and Deur \cite{Brodsky:2010ur} or  the dispersive approach of
Dokshitzer \emph{et al.} \cite{Dokshitzer:1995qm,Dokshitzer:1995zt}. A new review of the QCD running coupling in various approaches  can be found in Ref.
\cite{general}.

The remainder of this paper consists of the following sections. In Sec.~\ref{FRACTIONAL} we provide a brief discussion
of the theoretical formalism for different aspects of anQCD, including the $2\delta$anQCD model. In Sec.~\ref{pQCD} we deal with the
theoretical formalism of pQCD in DIS processes where the Jacobi polynomials  and the rational expansion for the moment of polarized nucleon structure function
are also presented. In Sec.~\ref{fapt} we employ the 2$\delta$anQCD model as the approach to evaluate the perturbation expansions and to extract the moments of
the polarized structure functions at the next-to-leading-orde (NLO) approximation. In Sec.~\ref{sec:sumrull} we also utilize this model to evaluate the Bjorken sum rule for polarized
nucleon structure functions. Finally we give our summary and conclusions in Sec.~\ref{summery}.

\section{{An overview of analytic QCD approaches}}\label{FRACTIONAL}
The anQCD approaches are intended to address the problems which were mentioned in the introduction, especially the second problem, i.e., the
presence of the Landau singularities in pQCD couplings $\alpha_s(Q^2)$. FAPT
\cite{ShS,MS,MSS,MSa,Sh1,Sh2,Karanikas:2001cs,Bakulev:2005gw,Bakulev:2006ex,Bakulev:2010gm,Bakulev:2008td} is one such approach. Other approaches
\cite{Brodsky:2010ur, Dokshitzer:1995qm,Dokshitzer:1995zt} could be applied, but we do not follow this line here. The calculations in the present work will be
performed using the $2\delta$anQCD approach \cite{CCEMAyala,Ayala:2014pha}, which has significant modifications with respect to FAPT.

For this purpose, we first give a brief description of (F)APT, a description essentially using Ref.~\cite{How}. After that we give a summarized description of
the $2\delta$anQCD model.

In the FAPT approach the running pQCD coupling {{}$a_s(Q^2) \equiv \alpha_s(Q^2)/4\pi$} of pQCD is transformed into a  holomorphic function of $Q^2$ in which first the
conversion  $a(Q^2) \mapsto \A_1(Q^2)$ is performed for the original pQCD coupling. The new coupling $\A_1(Q^2)$ is an analytic (holomorphic) function of $Q^2$
in the complex $Q^2$ plane with the only exception of the timelike semiaxis ($Q^2\leq 0$), and it is called the APT coupling constant.

As it will be seen later on, the APT coupling is given by a dispersion relation. In this relation there is the pQCD spectral density, denoted by $\rho^{\rm
(pt)}(\sigma) \equiv {\rm Im} \; \alpha_s(Q^2=-\sigma - i \epsilon)/\pi$ which in APT is not changed in the complex $Q^2$ plane for the whole negative axis,
given by the condition $\sigma \geq 0$. For the unphysical cut, i.e. $0 < Q^2 < \Lambda^2$, this spectral function is set equal to zero \cite{How}.
Following the same dispersion relation, based on the APT framework, the images (analogs) $\mathcal{A}_n(Q^2)$ of integer powers $a^n(Q^2)$ can also be
constructed.

In contrast with the original $a^n(Q^2)$ the couplings $\A_n(Q^2)$ are changing slowly at low $Q^2$  values. On the other hand, at large  $Q^2$  values these two
couplings are close to each other [$\A_1(Q^2) - a(Q^2) \sim \Lambda^2_{\rm QCD}/Q^2$].

The FAPT approach, which was summarized here, is an extension to noninteger powers $\nu$ in which $a^\nu$ maps to  $\A_\nu$
\cite{Karanikas:2001cs,Bakulev:2005gw,Bakulev:2006ex,Bakulev:2010gm,Bakulev:2008td}.
This case has many applications to DIS processes. One can refer to Ref. \cite{Sidorov2013} which contains  reasonable results in analyzing with the FAPT approach the
DIS data for the  hadron characteristics. Alongside the FAPT framework, there are various analytic QCD models which can be found in
Refs.~\cite{Nesterenko:1999np,Nesterenko:2004tg,Alekseev:2005he,Y.Srivastava,Webber:1998um,Cvetic:2006mk,Cvetic:2006gc,How,Prosperi:2006hx,Cvetic:2008bn}. One of
them is called the $2\delta$anQCD model \cite{CCEMAyala,Ayala:2014pha}, which will be discussed later on in this section.

In the framework the FAPT approach, one can obtain the following dispersion relation  for  the analogs (images) $\A_\nu^{(l)}$ of  the running coupling
$a^\nu(Q^2)\equiv\left(\alpha_s(Q^2)/\pi\right)^\nu $ at the $l$-loop order in the spacelike domain, by
applying the Cauchy theorem \cite{ShS,MS,MSS,MSa,Sh1,Sh2} :
\begin{equation}
 \A_\nu^{(l)}(L) =\int_0^{\infty} \frac{\rho_\nu^{(l)}[\sigma]}{\sigma+Q^2}d\sigma
 =
 \int_{-\infty}^{\infty} \frac{\rho_\nu^{(l)}(L_\sigma)}{1+{\rm exp}(L-L_\sigma)}dL_\sigma\;.
\label{AnuSD}
\end{equation}
In Eq.~(\ref{AnuSD}), $L_\sigma = \ln(\sigma/\Lambda^2)$ and $L=\rm{ln}(Q^2/\Lambda^2)$. The result of Eq.~(\ref{AnuSD})  does not contain any Landau
singularities.

The spectral function $\rho_\nu^{(l)}$ in Eq.~(\ref{AnuSD}) has the following representation \cite{ShS,MS,MSS,MSa,Sh1,Sh2}:
\begin{equation}{{}
\rho_\nu^{(l)}(L_\sigma) \equiv \frac{1}{\pi}{\rm Im} \left(a^\nu_{(l)}(L-i\pi)\right)
=\frac{{\rm sin}[\nu\varphi_{(l)}(L)]}{\pi\left( R_{(l)}(L) \right)^\nu}, \label{eq:rho}}
\end{equation}
where
\ba
R_{(l)}(L)= \big|a_{(l)}(L-i\pi) \big|,\nonumber\\
~~\varphi_{(l)}(L)=arg\left(a_{(l)}(L-i\pi)\right)\;. \label{eq:rhoF}
\ea
Considering Eq.~(\ref{eq:rho}) together with  Eq.~(\ref{AnuSD}), it is seen that the images $\A_\nu$ do not obey  the standard algebra such that  $\A_\nu\A_\mu
\neq \A_{\nu+\mu}$.

It can be proved that, at the one-loop approximation, $\varphi_{(1)}$ and $R_{(1)}$  in Eq.~(\ref{eq:rho}) are the simplest and are given, respectively, by \cite{How}

\begin{equation}
\varphi_{(1)}(L)={\rm arccos} \left(\frac{L}{\sqrt{L^2+\pi^2}} \right),
\quad
R_{(1)}(L)=\beta_0 \sqrt{L^2+\pi^2}.\label{eq:rho-1}
\end{equation}
One can reproduce at $Q^2=0$  the maximum value for $\A_1^{(1)}(L)$, i.e. $ \A_1^{(1)}(L=-\infty)$ which can be shown to have the well-known expression
\cite{ShS}
\begin{equation}
 \A_1^{(1)}(-\infty)=\int_{-\infty}^{\infty} \frac{dL_\sigma}{\beta_0(L_\sigma^2+\pi^2)} = \frac1{\beta_0} >  \A_1^{(1)}(L)\,.
\end{equation}
For this purpose, it is needed to substitute  Eq.~(\ref{eq:rho-1})  in Eq.~(\ref{eq:rho}) and then insert  in Eq.~(\ref{AnuSD}) the result  for $\rho_1^{(1)}$.

The situation becomes more complicated  when the two-loop approximation is considered. In this case, the solution of the QCD-$\beta$ function will yield  the
following result for  $a_{s(2)}$ in terms of the Lambert-$W$ function \cite{How}:
\be
{{}a_{s(2)}}=-\frac{1}{c_1}\frac{1}{1+W_{-1}(z_W(L))}\;.
\ee
{{}In this equation $z_W(L)$ is given by
\be z_W(L)=-c_1^{-1}
e^{-1-L \beta_0/c_1}\;,\ee
where
$c_1=\beta_1/\beta_0=(102-38 N_f/3)/(11-2 N_f/3)/4$ in which  $N_f$ is denoting the number of active quark flavors, and $Q^2=|Q^2| \exp(i \phi)$ is considered to have $0 \leq \phi < \pi$.}

Considering  Eq.(\ref{eq:rhoF}), the required expressions for $R_{(2)}$ and $\varphi_{(2)}$ would be, respectively,
\ba
R_{(2)}(L) &=& c_1({{}N_f)}\left| 1+W_{-1}(z_W (L+i\pi))\right|\;,
\nonumber\\
\varphi_{(2)}(L)&=& {\rm arccos}\left[\frac{{\rm Re}\left(1+W_{-1}(z_W (L+i\pi))\right)}{R_{(2)}(L)} \right]\;.
\ea
Here $W_{-1}(z)$ represents  the appropriate branch of the Lambert-$W$ function.
Extension of the calculations up to the four-loop  level can be found in Ref. ~\cite{Bakulev:2012sm}.

Because of the linearity of the transforms in Eq.~(\ref{AnuSD}) there is a one-to-one correspondence between the pQCD  and FAPT expansion  \cite{Sh1}.
This linearity can be exposed, considering single-scale quantity $D(Q^2,\mu^2_R)$  at the renormalization scale $\mu^2_R=Q^2$ within the minimal subtraction
scheme. In this regard the expansions for $D$ and its corresponding image, i.e., ${\cal D}$, have the following expansions \cite{How}:
\ba
\text{pQCD:}~D(Q^2)     &=&d_0~a^{\nu_0}(Q^2)~+ \sum_n d_n~a^{n+\nu_0}(Q^2)\;, \label{eq:DtoD}  \\
\text{FAPT:}~{\cal D}(Q^2)&=&d_0~\A_{\nu_0}(Q^2)+ \sum_n d_n~\A_{n+\nu_0}(Q^2) \label{eq:DtoD1}.
\ea
We note that $d_i$ coefficients are identical in both expansions, and the renormalization scale $\mu^2_R=Q^2$  is used.\\

After the summarized description of the FAPT approach, we now present briefly the other anQCD model, which at low-energy scales (where Landau singularities
appear in $a$) has a modified spectral function. It is therefore expected that, by applying this model, more reliable results are obtained, arising out of the
anQCD approach. This model, as we mentioned earlier, is called $2\delta$anQCD. Our calculations in this paper are entirely  based on utilizing this model.

In 2$\delta$anQCD \cite{CCEMAyala,Ayala:2014pha}, the spectral function $\rho(\sigma) \equiv {\rm Im} \mathcal{A}_1(-\sigma - i \epsilon)$ is equal to the
spectral function $\rho^{\rm (pt)}(\sigma) \equiv {\rm Im} a(-\sigma - i \epsilon)$ of the underlying pQCD coupling $a$ only for $\sigma \geq M_0^2$, where
$M_0^2 \sim 1 \ {\rm GeV}^2$ is regarded as the pQCD-onset scale {{}  because $\rho^{(2 \delta)}(\sigma)$ for $\sigma> M_0^2$ is taken equal to the discontinuity function $\rho^{\rm (pt)}(\sigma)$ of the underlying pQCD coupling $a_s(Q^2)$. The exact value of $M_0^2$  depends  on the chosen value for $c_2$  as the scheme-dependent parameter (for more delatils, see the Table 2 in Ref. \cite{Ayala:2014pha}).}

In 2$\delta$anQCD , in the low-$\sigma$ regime, i.e. $0 < \sigma < M_0^2$, the behavior of the
spectral function $\rho(\sigma) \equiv {\rm Im} \mathcal{A}_1(Q^2=-\sigma - i \epsilon)$ is controlled  by two positive delta functions. The coupling
$\mathcal{A}_1(Q^2)$  is presented by a dispersion integral which, in addition to the two delta functions, contains as well the $\rho^{\rm (pt)}(\sigma)$
function. The unknown parameters of the delta functions, and the $M_0$ scale, are determined by two requirements \cite{CCEMAyala,Ayala:2014pha,ana}. The first
one is to effectively match the model with the pQCD  at  large $|Q^2| >  \Lambda^2$. Second, the model should reproduce  the reported experimental rate for
the $\tau$ lepton semihadronic nonstrange $V+A$ decay  ratio, i.e.  $r_{\tau}^{\rm (D=0)} \approx 0.203$ .

The physical spacelike QCD quantities $D(Q^2)$, to be evaluated in anQCD with ${\A}_1$, are in practice, in general, represented in pQCD as a truncated
perturbations series of Eq.~(\ref{eq:DtoD}), and using the renormalization scale $\mu^2 = \kappa Q^2$ ($0 < \kappa \lesssim 1$)
{\small
\ba
{ D}(Q^2)^{[{N}]} &=&  a(\kappa Q^2)^{{\nu_0}} +
d_1(\kappa) a(\kappa Q^2)^{{\nu_0}+1} + \ldots\nonumber\\
&&+ d_{N-1}(\kappa) a(\kappa Q^2)^{{\nu_0}+{N}-1}.
\label{pt}
\ea
}
We point out that in anQCD the simple replacement $a(\kappa Q^2)^{{\nu_0}+m} \mapsto {\A}_1(\kappa Q^2)^{{\nu_0}+m}$ in Eq.~(\ref{pt}) is not applicable, since
it makes the perturbation series  diverge rapidly by increasing the power index $N$ (see \cite{GCtech}), and the nonperturbative contributions generated in this
way become erratic.
Therefore,
the formalism introduced in  Refs.~\cite{Cvetic:2006mk,Cvetic:2006gc} for the case of integer ${\nu_0}$  and the formalism in Ref.~\cite{ana} for the case of general real
$\nu_0$, are needed to overcome the difficulty, in any anQCD. Consequently, the following replacement takes place in Eq.~(\ref{pt}) in anQCD:
{\small
\be
a(\kappa Q^2)^{{\nu_0}+m} \mapsto
{\A}_{{\nu_0}+m}(\kappa Q^2)
\quad \left[ \not= {\A}(\kappa Q^2)^{{\nu_0}+m} \right] \;.
\label{an1}
\ee
}
The holomorphic power analog ${\A}_{{\nu_0}+m}(Q^2)$
can be obtained from ${\A}_1(Q^2)$ by the following construction.
First one needs the logarithmic derivatives of ${\A}_1(Q^2)$ which are given by \cite{ana}
{\small
\be
{\tilde{\mathcal{A}}}_{n+1}(Q^2) \equiv \frac{(-1)^n}{\beta_0^n n!}
\left( \frac{ \partial}{\partial \ln Q^2} \right)^n
{\A}_1(Q^2) \ , \qquad (n=0,1,2,\ldots) \ ,
\label{tAn}
\ee
}
where  $\beta_0=(11-2 N_f/3)/4$, and it is obvious that ${\tA}_1 \equiv {\A}_1$.

Considering  anQCD coupling ${\A}_1$,  Eq.~(\ref{tAn}) can be written in terms of the discontinuity function ${\rho}(\sigma) \equiv {\rm Im} {\A}_1(-\sigma-i
\epsilon)$   where the Cauchy theorem is used  along the cut line. Therefore, Eq.~(\ref{tAn}) can be written as follows \cite{Cvetic:2006mk,Cvetic:2006gc,ana}:
{\small
\be
{\tA}_{n+1}(Q^2) = \frac{1}{\pi} \frac{(-1)}{\beta_0^n \Gamma(n+1)}
\int_{0}^{\infty} \ \frac{d \sigma}{\sigma} {\rho}(\sigma)
{\rm {Li}}_{-n} ( -\sigma/Q^2 ) \ .
\label{disptAn2}
\ee
}
{{}Here $\Gamma(n)$ is denoting  the usual Gamma function and $\rm {Li}_{-n}$ is defined below in the general case  by Eq.(\ref{Li_nugen}).}

The generalization to the noninteger $n \mapsto {\nu}$ is given by \cite{ana}
{\small
\be
{\tA}_{{\nu}+1}(Q^2) = \frac{1}{\pi} \frac{(-1)}{\beta_0^{{\nu}} \Gamma({\nu}+1)}
\int_{0}^{\infty} \ \frac{d \sigma}{\sigma} {\rho}(\sigma)
{\rm {Li}}_{-{\nu}}\left( - \frac{\sigma}{Q^2} \right) \quad (-1 < {\nu}) \ .
\label{tAnu1}
\ee
}
{{}Instead of  ${\rho}$} the above quantity can be expressed in terms of ${\A}_1$ ($\equiv \tA_1$) \cite{ana}:
{\small
\ba
{\tilde{\mathcal{A}}}_{{\delta}+m}(Q^2) & = &
K_{{\delta}, m}
\left(\frac{d}{d \ln Q^2}\right)^{m}
\int_0^1 \frac{d \xi}{\xi} {\A}_1(Q^2/\xi) \ln^{-{\delta}}\left(\frac{1}{\xi}\right) \ .\nonumber\\
\label{tAnu2}
\ea
}
Here $K_{{\delta},m} =(-1)^m \beta_0^{-{\delta}-m+1}/[\Gamma({\delta}+m)\Gamma(1-{\delta})]$ in which $m=0,1,2,\ldots$ and $0\leq {{\delta}} <1$.
Equation~(\ref{tAnu2}) originates  from Eq.~(\ref{tAnu1}), using the following  expression for the ${\rm Li}_{-\nu}(z)$ function
\cite{KKSh}:
{\small
\ba
{\rm {Li}}_{-{n}-{\delta}}(z) &=& \left( \frac{d}{d \ln z} \right)^{{n}+1}
\left[ \frac{z}{\Gamma(1 - {\delta}) } \int_0^1 \frac{d \xi}{1 - z \xi}
\ln^{-{\delta}} \left( \frac{1}{\xi} \right) \right]
\nonumber\\
&&\quad ({n} =-1, 0, 1, \ldots; 0 < {{\delta}} < 1) \ .
\label{Li_nugen}
\ea
}
Combining various generalized logarithmic derivatives, the analytic analogs ${\A}_{{\nu}}$ of powers $a^{{\nu}}$  are obtained \cite{ana}

\be
{\A}_{\nu}={\tA}_{{\nu}}
+\sum_{m\geq1} {\tk}_m({\nu}) {\tA}_{{\nu}+m} \ ,
\label{AnutAnu}
\ee
where the coefficients ${\tk}_m({\nu})$ were obtained in Ref.~\cite{ana} for general $\nu$. \textcolor{black}{It can be shown that this approach, in the specific case
of APT ($\rho(\sigma)=\rho^{\rm (pt)}(\sigma)$; $\sigma \geq 0$), gives the same result as the expression  ${\A}_{\nu}^{\rm (FAPT)}(Q^2)$ in Eq.~(\ref{AnuSD}).
However, the approach Eqs.~(\ref{tAnu1})-(\ref{AnutAnu}) is applicable in any anQCD.}
We recall that in the $2\delta$anQCD model
 at high momenta, $\sigma \geq {M_0^2}$,  the discontinuity function  $ {{\rho}}(\sigma) \equiv {\rm Im}  {\A}_1(-\sigma - i \epsilon)$ is given
by  ${\rho^{\rm (pt)}}(\sigma)$ [$\equiv {\rm Im} \  a(-\sigma - i \epsilon)$]. In the  low-momentum region it is parametrized  by two delta functions, such that
\cite{CCEMAyala,Ayala:2014pha}
{\small
\be
{\rho}^{({2 \delta})}(\sigma) =  \pi {F_1^2} \delta(\sigma - {M_1^2})
+ \pi {F_2^2} \delta(\sigma - {M_2^2}) + \Theta(\sigma-{M_0^2}) {\rho}^{\rm (pt)}(\sigma)\;.
\label{rho2d}
\ee
}
{{}where $\Theta$ is the step (Heaviside) function.}
Therefore the 2$\delta$anQCD coupling is represented by
\ba
\lefteqn{
{\tA}_{{\nu}}^{({2 \delta})}(Q^2) =
\frac{(-1)}{\beta_0^{\nu} \Gamma(\nu\!+\!1)} {\bigg\{}
\sum_{{j=1}}^{{2}}
\frac{{F_j^2}}{{M_j^2}}
{\rm {Li}}_{-\nu}\left( - \frac{{M_j^2}}{Q^2} \right)
}
\nonumber\\
&&
+ \frac{1}{\pi}
\int_{{M_0^2}}^{\infty} \ \frac{d \sigma}{\sigma} {\rm Im} a
(-\sigma\!-\!i \epsilon)
{\rm {Li}}_{-\nu}\left( - \frac{\sigma}{Q^2} \right) {\bigg\}}.
\label{tAnu2d}
\ea
In Eq.~(\ref{tAnu2d})  the parameters ${F_j^2}$ and ${M_j}$ ($j=1,2$)
are determined by the requirement that the deviation from the conventional pQCD result at high ${Q^2} > \Lambda^2$ is such that  ${\A}_{{\nu}}^{({2
\delta})}(Q^2) - {a(Q^2)}^{{\nu}}  \sim ( {\Lambda^2}/{{Q^2}} )^{5}$, i.e., essentially zero [in (F)APT this deviation is $\sim  ( {\Lambda^2}/{{Q^2}} )^{1}$].
These are, in fact, four requirements. The pQCD-onset scale $M_0$ is fixed by the condition that the model should reproduce the measured semihadronic decay ratio
${r_{\tau} \approx 0.203}$ for the strangeless and massless $V+A$ tau lepton \cite{CCEMAyala,Ayala:2014pha,AC}.

In the 2$\delta$anQCD model the coupling ${ a}$ of the underlying pQCD is chosen for convenience in terms of the Lambert-$W$ function (which \textit{Mathematica} can
evaluate without problems) such that

{\small
\be
a(Q^2)=-\frac{1}{c_1}\frac{1}{\left[ 1-c_2/c_1^2+W_{\mp1}(z_{\pm}) \right]} \ .
\label{aptexact}
\ee
}
As defined before, in this equation  $c_1=\beta_1/\beta_0$, $c_2 = \beta_2/\beta_0$, and  $Q^2=|Q^2|{\rm e}^{i\phi}$ {{} where $\phi$ is the required argument in which one can follow the calculations in the complex $Q^2$ plane}. The two universal beta coefficients are $\beta_0=(11 - 2
N_f/3)/4$ and $\beta_1 = (102 - 38 N_f/3)/16$.
The upper and lower sign refer to  $\phi \geq 0$  and $\phi < 0$, respectively, and the $z$ variable is
{\small
\be
z_{\pm}=(c_1{\rm e})^{-1} ({|Q^2|}/{\Lambda_L^2} )^{-\beta_0/c_1}
{\rm exp}\left[i (\pm\pi-{\beta_0} \phi/{c_1} ) \right].
\label{zina2l}
\ee }
{{}The coupling (\ref{aptexact}) is in a renormalization scheme called the Lambert scheme. In these  schemes, $c_n (\equiv \beta_n/\beta_0)$  $= c_2^{n-1}/c_1^{n-2}$  (n=3,4,$\cdots$), i.e., all the higher $\beta_n$ scheme parameters (n=3,4,$\cdots$) are fixed by $c_2$ \cite{Ayala:2014pha}. The acceptable values of the scheme parameter $c_2$ can be in  the interval $-5.6 < c_2 < -2$, with $c_2=-4.9$ as the  preferred (and chosen) value  \cite{Ayala:2014pha}.}
The Lambert scale $\Lambda_L= 0.255$ GeV is determined in such a way that it corresponds in the ${\overline {\rm MS}}$ scheme at {{} $\mu^2=k,\; Q^2=M_Z^2$ (with $N_f=5$ and $k=1$, see the text below Eq.(\ref{eq:sumrulfapt}))} to the value $\alpha_s(M_Z^2;{\overline {\rm MS}})=0.1184$. \textcolor{black}{The scaling in the pQCD coupling (\ref{aptexact}) corresponds to the ${\overline {\rm MS}}$ scaling, i.e., $a(Q^2) - a^{\overline {\rm MS}}(Q^2) \sim a^3$.}

\begin{figure}[!htb]
	\vspace*{0.5cm}
\includegraphics[clip,width=0.45\textwidth]{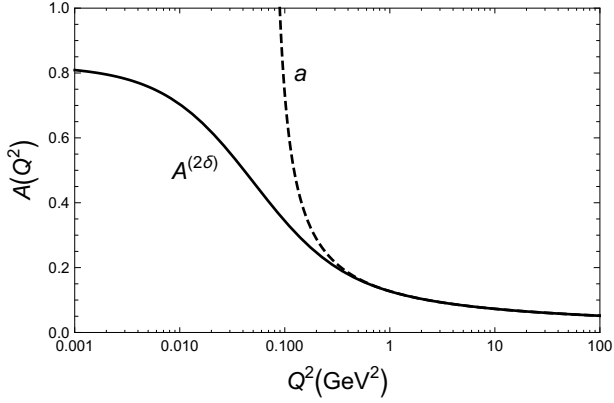}
	\begin{center}
	  \caption{\small The coupling ${\cal A}_1^{(2 \delta)}$ in $2\delta$anQCD, with $N_f=3$, as a function of $Q^2$. \textcolor{black}{The dashed line
represents the underlying pQCD coupling Eq.~(\ref{aptexact}), with $c_2=-4.9$.}}
\label{fig:A-1}
	\end{center}
\end{figure}

In Fig.~\ref{fig:A-1} the 2$\delta$anQCD {{} running strong  coupling} is plotted \textcolor{black}{as a function  of $Q^2$}, and compared with the coupling of the
underlying pQCD. As can be seen, the 2$\delta$anQCD coupling ${\A}_1$ has a finite value when $Q^2 \to 0$, while the coupling $a$ from the underlying pQCD
increases rapidly and goes to infinity near the Landau branching point $Q^2=\Lambda^2$.

Keeping in mind Eq.~(\ref{AnutAnu}), any QCD observable can be calculated within $2\delta$anQCD (or any anQCD). In the following sections we employ this model to
analyze the polarized nucleon structure function and the related Bjorken sum rule.

\section{NLO evolution of polarized parton densities, Rational approximation and  Jacobi polynomials}\label{pQCD}

We evaluate here the polarized nucleon structure function.
The precision depends on the method to calculate the structure function.
It is therefore necessary to characterize the employed method.
One of the reliable methods is the use of the Jacobi polynomials which leads to sufficient precision in the evolution of the function with $Q^2$.
\textcolor{black}{Alternatively, using the inverse Mellin transformation, the result in Bjorken $x$ space can be directly obtained} from the  evolved  moments of the structure functions  in $n$ space.

Before employing the Jacobi method, we present some technical details about the evolution of the polarized parton densities in the moment $n$ space. In this
context, an approximate method (rational approximation) is introduced which yields required expansions for the considered moments.
For this purpose, we need  $\Delta P^{(0)}_{NS}(n)$  which is the $n${th} moment of
the nonsinglet splitting function at the leading order~\cite{Lampe:1998eu}:
\begin{eqnarray}\label{eq:splitingLO}
\Delta {P^{(0)}_{NS}(n)}&=&\frac{4}{3}\left[\frac{3}{2} +\frac{1}{n(n+1)}-
2 S_1(n)\right]\;.
\end{eqnarray}
In this equation $S_1(n)\equiv\sum\limits^n_{j=1} 1/j=\psi (n+1)+\gamma_E$, where $\psi(n)\equiv \Gamma'(n)/\Gamma(n)$, and $\gamma_E=0.577216$ is the
Euler-Mascheroni constant.

{{{}{As we specify later on, our calculations  throughout the whole manuscript  are based on using the 2$\delta$anQCD model alongside the underlying pQCD as represented by the (pQCD) coupling Eq.~(\ref{aptexact}). They are both in the specific Lambert ($c_2=-4.9$) scheme, but in the rest of this section we use terms and phrases which are valid in general schemes. In the next section, which contains the main results for the nucleon structure function $g_1$, the required expressions are used in the Lambert scheme, but here a general discussion is followed. In this regard we resort to a solution of the renormalization group equation for the pQCD coupling in NLO approximation that is given in Ref. \cite{larin}. Then the solution of the NLO evolution equation for the moments of the structure function leads to }} ~\cite{YndBook}:
\begin{eqnarray}
&&\Delta M_\text{NS}(n,Q^2) =
\nonumber\\
&&\frac{1+\Delta C_\text{NS}^{(1)}(n) a_s(Q^2)}
{1+\Delta C_\text{NS}^{(1)}(n) a_s(Q_0^2)} \left(\frac{1+(b_1/b_0) a_s(Q^2)}
{1+(b_1/b_0) a_s(Q_0^2)} \right)^{\Delta p(n)}
\nonumber\\
&& \times \left[\frac{a_s(Q^2)}{a_s(Q_0^2)}\right]^{\Delta d_\text{NS}(n)}\Delta M_\text{NS}(n,Q_0^2)\;.
 \label{eq:mns}
\end{eqnarray}
In the above equation the Wilson coefficient $\Delta C^{(1)}_{NS}$ is given by  \cite{Lampe:1998eu}

\begin{eqnarray}
\Delta C^{(1)}_{NS}&=& \frac{4}{3}\biggl[-S_2(n)+(S_1(n))^2
+\left(\frac{3}{2}-
\frac{1}{n(n+1)}\right)\nonumber\\
&&\times S_1(n)
+\frac{1}{n^2}
+\frac{1}{2n}+\frac{1}{n+1}-\frac{9}{2} \biggr]
\label{eq:wilsonnlo}
\end{eqnarray}
where $S_2(n)\equiv\sum^n_{j=1} 1/j^2=\pi^2/6 -\psi '(n+1)$
in which $\psi '(n)=(d/dn)^2 \ln \Gamma (n)$.

In Eq.~(\ref{eq:mns}),  $\Delta M_\text{NS}(n,Q_0^2)$ can be obtained from the moment of the polarized parton densities, $\Delta f_\text{NS}(n,Q_0^2)$, as
follows:
\begin{equation}\label{eq:mnsQ0}
\Delta M_\text{NS}(n,Q_0^2)=\left(1+\Delta C_\text{NS}^{(1)}(n) a_s(Q_0^2) \right)\Delta f_\text{NS}(n,Q_0^2)\, .
\end{equation}
$\Delta f_\text{NS}(n,Q_0^2)$, the moment of the parton densities at the initial scale $Q_0^2$,
\textcolor{black}{will be discussed later on in this section.}
The functions $\Delta d_\text{NS}(n)$ and $\Delta p(n)$ in Eq.~(\ref{eq:mns})
\textcolor{black}{are the moments of the splitting functions and}
are given by, respectively, by~\cite{YndBook}
\begin{eqnarray}\label{eq:mnssplitinging}
\Delta d_\text{NS}(n)&=&-2\Delta P_\text{NS}^{(0)}(n)/b_0,\\
\Delta p(n) &=& -4 \left(\frac{\Delta P_\text{NS}^{(1)}(n)}{b_1}-\frac{\Delta P_\text{NS}^{(0)}(n)}{2b_0}
\right)\,.\label{eq:mnsspliting}
\end{eqnarray}
Here, the explicit expression for $\Delta P^{(1)}_{NS\pm}$ is required which is presented in the Appendix.

One of the methods to obtain a solution for the moments of the structure function in Eq.~(\ref{eq:mns}) is to use the rational approximation, which is
indispensable when we employ the analytic perturbation theory. On this basis, the following factor in Eq.~(\ref{eq:mns}):
\begin{equation}
\left(\frac{1+(b_1/b_0) a_s(Q^2)}
{1+(b_1/b_0) a_s(Q_0^2)} \right)^{\Delta p(n)}\;,
\label{eq:approxM}
\end{equation}\\
which is denoted by $\Delta m(n,Q^2)$ has the following expansions up to $\mathcal{O}(a)$ and
$\mathcal{O}(a^2)$ accuracy,  respectively~\cite{YndBook}:
\begin{widetext}
\begin{eqnarray}
\Delta m^{(1)}_\text{pQCD}(n,Q^2)&\simeq& \frac{1+(b_1/b_0)\Delta p(n) a_s(Q^2)}
{1+(b_1/b_0)\Delta p(n) a_s(Q_0^2)},\nonumber \\
\Delta m^{(2)}_\text{pQCD}(n,Q^2)&\simeq& \frac{1+(b_1/b_0)\Delta p(n)
a_s(Q^2)+(b_1^2/2b_0^2)\Delta p(n)(\Delta p(n)-1)a_s^2(Q^2)}
{1+(b_1/b_0)\Delta p(n) a_s(Q_0^2)+(b_1^2/2b_0^2)\Delta p(n)(\Delta p(n)-1)a_s^2(Q_0^2)}\;.
\label{eq:mellin}
\end{eqnarray}
\end{widetext}
The polarized splitting function $\Delta p(n)$ in Eq.~(\ref{eq:mellin}) is obtained by  Eq.~(\ref{eq:mnsspliting}).

To construct the moments of structure functions, one needs the polarized parton densities at initial $Q_0^2$ as the  input densities. We take them from
the data-based KATAO PDFs  at $Q^2_0=4 \ {\rm GeV}^2$  such that ~\cite{Khorramian:2010qa}
\begin{eqnarray}\label{eq:parametrizationsQ0}
x \Delta u_v(x, Q_0^2) &=& {\cal N}_{u_v} \, \eta_{u_v} \, x^{\alpha_{u_v}} (1 - x)^{\beta_{u_v}} \, (1 + \gamma_{u_v} x) \,,
\nonumber\\
x \Delta d_v(x, Q_0^2) &=& {\cal N}_{d_v} \, \eta_{d_v} \, x^{\alpha_{d_v}} (1 - x)^{\beta_{d_v}} \, (1 + \gamma_{d_v} x).
\end{eqnarray}
\textcolor{black}{One of the advantages of the value $Q^2_0=4 \ {\rm GeV}^2$ here is that the $2\delta$anQCD coupling at this scale practically coincides with the
underlying pQCD coupling (\ref{aptexact}).}
The numerical  values for $\eta_{q_v}$ and $\alpha_{q_v},\beta_{q_v}$ and $\gamma_{q_v}$ parameters are listed in Table~\ref{tab:fit}. The normalization constants
${\cal N}_{q_v}$,
\begin{equation}\label{eq:normal}
\frac{1}{{\cal N}_{q_v}}=\bigg(1+\gamma_{q_v}\frac{\alpha_{q_v}}{\alpha_{q_v}+\beta_{q_v}+1}\bigg)B(\alpha_{q_v},\beta_{q_v}+1),
\end{equation}
are chosen such that $\eta_{q_v}$ is considered as the first moment of $x \Delta q_v(x, Q_0^2)$, i.e.  $\eta_{q_v}=\int_{0}^{1}dx\Delta q_v(x, Q_0^2)$.
The $B(a,b)$ in Eq.~(\ref{eq:normal}) is the Euler beta function. Details of computations to obtain  numerical values for the first moments of  $\Delta u_{v}$
and $\Delta d_{v}$ are presented in the following subsection. {{} At the NLO approximation, the results of calculations are scheme independent ($c_2$ independent), and we can use the Lambert scheme.}

\begin{table}
\begin{tabular}{>{\centering}p{0.3in}>{\centering}p{0.3in}c>{\centering}p{0.3in}>{\centering}p{0.3in}c}
\hline\hline
 & $\eta_{u_{v}}$  & $~0.928\ (fixed)~$  &  & $\eta_{d_{v}}$  & $-0.342\ (fixed)$ \tabularnewline
$\Delta u_{v}$  & $\alpha_{u_{v}}$  & $0.535\pm0.022$  & $\Delta d_{v}$  & $\alpha_{d_{v}}$  & $0.530\pm0.067$ \tabularnewline
 & $\beta_{u_{v}}$  & $3.222\pm0.085$  &  & $\beta_{d_{v}}$  & $3.878\pm0.451$ \tabularnewline
 & $\gamma_{u_{v}}$  & $8.180\ (fixed)$  &  & $\gamma_{d_{v}}$  & $~4.789\ (fixed)~$ \tabularnewline
\hline\hline
\end{tabular}
\caption{{\small  Numerical values for the  first moment parameters of $\Delta u_{v}$ and $\Delta d_{v}$ and their statistical errors at the input scale $Q_{0}^{2}=4$ GeV$^{2}$ in the NLO approximation, based on the KATAO parameterization model
~\cite{Khorramian:2010qa}.}
\label{tab:fit}}
\end{table}
\subsection{First moments of $\Delta u_{v}$ and $\Delta d_{v}$}
The parameters  $\eta_{u_{v}}$ and $\eta_{d_{v}}$ are the first moments
of the  polarized valence quark densities, $\Delta u_{v}$ and $\Delta d_{v}$,  respectively.
These moments can be related to $F$ and $D$ quantities \cite{Tanabashi:2018oca} as measured in neutron and hyperon $\beta$-decays

\begin{eqnarray}\label{eq:q2q8}
a_{3} & = & \int_{0}^{1}dx\:\Delta q_{3}=\eta_{u_{v}}-\eta_{d_{v}}=F+D\ ,\\
a_{8} & = & \int_{0}^{1}dx\:\Delta q_{8}=\eta_{u_{v}}+\eta_{d_{v}}=3F-D\ .
\end{eqnarray}
In these equations  $a_{3}$ and $a_{8}$ are the nonsinglet combinations of the first
moments which are constructed from  the polarized parton densities such that
\begin{eqnarray}\label{eq:q2q8parton}
\Delta q_{3} & = & (\Delta u+\Delta\overline{u})-(\Delta d+\Delta\overline{d})\ ,\\
\Delta q_{8} & = & (\Delta u+\Delta\overline{u})+(\Delta d+\Delta\overline{d})-2(\Delta s+\Delta\overline{s})\,.
\end{eqnarray}

\noindent Doing a reanalysis for $F$ and $D$ with updated {{}$\beta$-decay}
constants, one obtains  $F=0.464\pm0.008$ and $D=0.806\pm0.008$ ~\cite{Tanabashi:2018oca}.
{{} Based on Eq.(\ref{eq:q2q8}) and  considering the  experimental values for $F$ and $D$,}  the following numerical values are obtained for the first moments of the polarized valence densities:

\begin{eqnarray}\label{eqetauvetadv}
\eta_{u_{v}} & = & +0.928\pm0.014\ ,\label{eq:etauv}\\
\eta_{d_{v}} & = & -0.342\pm0.018\ .\label{eq:etadv}
\end{eqnarray}

Full results for the moments of the nucleon  polarized structure functions at the NLO approximation are now accessible. It is therefore necessary to obtain the structure functions in
Bjorken-$x$ space. This can be done using the Jacobi polynomial method, which we summarize in the next subsection.
\subsection{The Jacobi polynomial method}
Based on the Jacobi polynomial method,the polarized structure function, $xg_1^{(NS)}(x,Q^2)=xg_1^{p}(x,Q^2)-xg_1^{n}(x,Q^2)$, can be expanded as
follows~\cite{Khorramian:2010qa,Kataev:1994rj,Khanpour:2017fey,Khanpour:2017cha,Taghavi-Shahri:2018ege}:
\begin{equation}
xg_{1}^{NS}(x,Q^{2})=x^{\beta}(1-x)^{\alpha}\ \sum_{n=0}^{N_{max}}a_{n}(Q^{2})\ \Theta_{n}^{\alpha,\beta}(x)\ .\label{eq:xg1}
\end{equation}
In the above equation, $\Theta_{n}^{\alpha,\beta}(x)$ are Jacobi polynomials of order $n$, and $N_{max}$ is the maximum order of expansion. Jacobi polynomials
provide a method to separate  the main
part of the $x$ dependence of the structure function into the weight
function $x^{\beta}(1-x)^{\alpha}$, while the $Q^{2}$ dependence is contained in
the Jacobi moments $a_{n}(Q^{2})$  ~\cite{Parisi:1978jv}.

Jacobi polynomials fulfill an orthogonality relation
\begin{equation}
\int_{0}^{1}dx\; x^{\beta}(1-x)^{\alpha}\Theta_{k}^{\alpha,\beta}(x)\Theta_{l}^{\alpha,\beta}(x)=\delta_{k,l}\ .\label{eq:ortho}
\end{equation}
Using this, Eq.~(\ref{eq:xg1}) can be inverted to yield the Jacobi moments $a_{n}(Q^{2})$:
\begin{eqnarray}
a_{n}(Q^{2}) & = & \int_{0}^{1}dx\; xg_{1}(x,Q^{2})\Theta_{n}^{\alpha,\beta}(x)\nonumber \\
 & = & \sum_{j=0}^{n}c_{j}^{(n)}(\alpha,\beta)\ { \Delta M}[xg_{1}^{NS},j+2]~\ .\label{eq:aMom}
\end{eqnarray}
To derive the last line in the above equation, it is needed to substitute  Eq.~(\ref{eq:xg1})
for $xg_{1}^{NS}(x,Q^{2})$ into the first line of Eq.~(\ref{eq:aMom}) and to use the Mellin transform
\begin{eqnarray}
{{\Delta M}}[xg_{1}^{NS},N] & \equiv & \int_{0}^{1}dx\ x^{N-2}\ xg_{1}^{NS}(x,Q^{2})\ .\label{eq:Mellin2}
\end{eqnarray}
The polarized structure function $xg_{1}^{NS}(x,Q^{2})$ can now be related to Mellin moments as follows \cite{Khorramian:2010qa}:
\begin{eqnarray}
xg_{1}^{NS}(x,Q^{2}) & = & x^{\beta}(1-x)^{\alpha}\sum_{n=0}^{N_{max}}\Theta_{n}^{\alpha,\beta}(x)\nonumber \\
 & \times & \sum_{j=0}^{n}c_{j}^{(n)}{(\alpha,\beta)}\ { {\Delta M}}[xg_{1}^{NS}, j+2]\ .\label{eq:eg1Jacob}
 \end{eqnarray}
It is required to choose the  set $\{N_{max},\alpha,\beta\}$ such that an optimal convergence of the series is achieved. This convergence should contain  the
whole kinematic region and cover the related experimental data. An improvement is achieved  for $\alpha=3.0$, and $\beta=0.5$  while $N_{max}$ varies between
7 and 9 \cite{Taghavi-Shahri:2018ege}.

\textcolor{black}{As an alternative to the above method,}
the structure function  $xg_{1}^{NS}(x,Q^{2})$ may be obtained by using the inverse  Mellin transform  for the moments of the structure function. On this basis,
one obtains for $xg_{1}^{NS}(x,Q^{2})$ the following expression in which a convenient path of integration is chosen:
\begin{eqnarray}\label{eq:inversmellin}
 xg_{1}^{NS}(x,Q^{2})&=&\frac{1}{\pi}\int_{0}^{5+\frac{10}{ln(1/x)}}dz{\text Im}[e^{i\phi}x^{1-c-zexp(i\phi)}\nonumber\\&&{\Delta M}_{NS}(N=c+ze^{i\phi})].
 \end{eqnarray}
In this integration, it is assumed  $c=1.9$ and $\phi=\frac{3\pi}{4}$~\cite{Vogt:2004ns}.

Now we are able to construct the polarized structure function in the 2$\delta$anQCD model, based on the Jacobi polynomials which is considered in the following
section.
\section{Extracting the polarized structure function, using the 2$\delta$anQCD model}\label{fapt}
Here we are going to employ the 2$\delta$anQCD model to construct the polarized nucleon structure function. {{} This model, like the underlying pQCD based on Eq.(\ref{aptexact}) as the coupling constant of series expansions, is considered in the Lambert scheme.}

\textcolor{black}{In the anQCD approach, we replace the power $a_s^{\nu} \equiv a^{\nu}/4^{\nu}$ of the conventional pQCD by the analytic coupling ${\cal
A}_{\nu}/4^{\nu}$. Note that $\A_{\nu}(Q^2)$ is the analog (image) of $a(Q^2)^{\nu} \equiv (\alpha_s(Q^2)/\pi)^{\nu}$, and $a_s(Q^2)=a(Q^2)/4$.}
In this way, the moments $\Delta \mathcal{M}_\text{NS}$ as the analytic images of $ \Delta M_\text{NS}$ are obtained. We recall that $\nu$ in the analytic
coupling  ${\cal A}_{\nu}$ is an expansion order index rather than a power index.
\textcolor{black}{For simplicity, we will denote in this section}
\begin{equation}
A_{s,\nu}(Q^2) \equiv \frac{{\A}_{\nu}(Q^2)}{4^{\nu}},
  \label{AvsmathA}
\end{equation}
i.e., $A_{s,\nu}$ is the anQCD image (analog) of $a_s^{\nu}$, while ${\A}_{\nu}$ is (always) the anQCD image (analog) of $a^{\nu}$.
When the first rational approximation in Eq.~(~\ref{eq:mellin}) is employed, the mentioned replacement in Eq.~(\ref{eq:mns}) leads to the following result
{\cite{How}}:
\begin{eqnarray}
  &&\Delta {\cal M}_\text{NS}(n,Q^2)=
  \nonumber\\&&
  \frac{ A_{s,\Delta d_\text{NS}(n)}(Q^2)+
\left(\Delta C_\text{NS}^{(1)}(n)+ \frac{b_1}{b_0} \Delta p(n) \right) A_{s,\Delta d_\text{NS}(n)+1}(Q^2)}
{ A_{s,\Delta d_\text{NS}(n)}(Q_0^2) +
\left(\Delta C_\text{NS}^{(1)}(n)+ \frac{b_1}{b_0} \Delta p(n) \right) A_{s,\Delta d_\text{NS}(n)+1}(Q_0^2) }
\nonumber\\&& \times {\Delta\cal M}_\text{NS}(n,Q_0^2)\;.
 \label{eq:mnsapt}
\end{eqnarray}
We then use this formalism, and evaluate $A_{s,\nu}$ with the help of the \textit{Mathematica} package that is called anQCD.m as introduced in Ref. ~\cite{Ayala:2014pha}.

Using the corresponding command for the analytic coupling ${\A}_{\nu}$ in the $2\delta$anQCD model in the two-loop approximation, we write ~\cite{Ayala:2014pha}
\begin{equation}\label{eq:anu}
  A_{s,\nu}(Q2)=\frac{A2d2l[3, 0,\nu, Q2, 0]}{4^{\nu}}\;.
\end{equation}
Substituting Eq.~(\ref{eq:anu}) in Eq.~(\ref{eq:mnsapt}),  with $Q2=Q^2>0$, leads to the numerical result for the moments of the structure function in the anQCD
approach. We note that there are some other commands in Ref.~\cite{Ayala:2014pha} for the anQCD coupling in 2$\delta$anQCD model but the one in Eq.~(\ref{eq:anu}) is
more appropriate for our numerical purpose with sufficient precision. \textcolor{black}{Namely, ${\A}_{\nu}(Q^2)=A2d2l[3,0,\nu,Q2,0]$, for positive $Q^2=Q2>0$ and
with $N_f=3$, is constructed as the truncated sum of two terms in Eq.~(\ref{AnutAnu}), i.e., ${\A}_{\nu} = {\tA}_{\nu} + \tk_1(\nu) {\tA}_{\nu+1}$, which is
sufficient for our NLO analysis.}

\begin{figure}[H]
	\vspace*{0.5cm}
	\includegraphics[clip,width=0.41\textwidth]{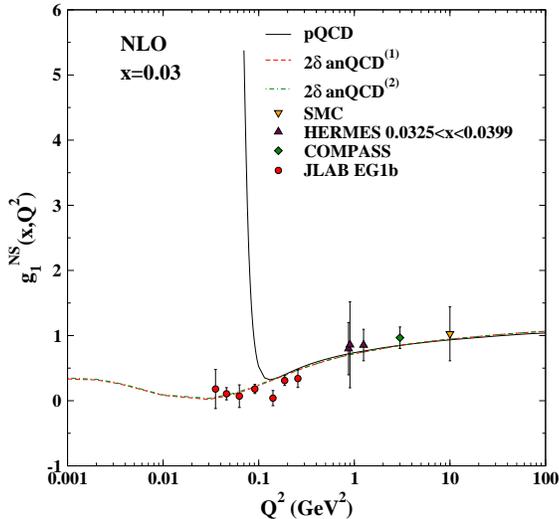}
	\vspace{-0.6cm}
	\begin{center}
		\caption{\small (Color online) Nonsinglet structure function $g_1^{(NS)}(x,Q^2)$ as a function of $Q^2$ at NLO. The dashed lines represent the
$2{\delta}$anQCD results, using   $\Delta { {m}}^{(1)}_\text{anQCD}$ and $\Delta { {m}}^{(2)}_\text{anQCD}$ in Eqs.(\ref{eq:mfapt} and \ref{eq:mfapt1}). The solid line is the underlying pQCD one.A comparison with the available experimental data~\cite{Airapetian:2006vy,Adeva:1998vv,Adolph:2015saz,priv} has also been done.
\label{fig:xgnsQdependent}}
	\end{center}
\end{figure}

As more explanation, it should be said that $ \texttt{A2d}N\texttt{l}[N_f,n,\nu,|Q^2|,\phi]$ is representing the $N$-loop analytic 2$\delta$anQCD coupling
${\cal A}_{n+\nu}^{(2\delta)}(Q^2, N_f)$ of fractional power $n+\nu$ where $\nu > -1$  and index $n$ is such that $n=0,1,\ldots,N-1$. Here the active quark
flavor $N_f$ is fixed. In the Euclidean domain the energy scale $Q^2$ is given by $Q^2 = |Q^2| \exp(i \phi)\in {\mathcal C} \backslash [-M^2_{\rm
thr.},-\infty)$ in which $M_{\rm thr.}^2 = M_2^2$ that is applicable  for the ${\rm N}^{n-1}{\rm LO}$ truncation approach~\cite{Ayala:2014pha}.

It is evident that the anQCD analog of Eq.~(\ref{eq:mellin}) is
\begin{widetext}
\begin{eqnarray}\label{eq:mfapt}
\Delta { {\rm m}}^{(1)}_\text{anQCD}(n,Q^2)&\simeq& \frac{1+(b_1/b_0) \Delta p(n) A_{s,1}(Q^2)}
{1+(b_1/b_0)\Delta p(n) A_{s,1}(Q_0^2)}, \\
\Delta {\rm m}^{(2)}_\text{anQCD}(n,Q^2)&\simeq& \frac{1+(b_1/b_0)\Delta p(n)
A_{s,1}(Q^2)+(b_1^2/2 b_0^2)\Delta p(n)(\Delta p(n)-1) A_{s,2}(Q^2)}
{1+(b_1/b_0)\Delta p(n) A_{s,1}(Q_0^2)+(b_1^2/2 b_0^2)\Delta p(n)(\Delta p(n)-1) A_{s,2}(Q_0^2)}.\label{eq:mfapt1}
\end{eqnarray}
\end{widetext}

One can construct from Eqs. (\ref{eq:mellin}) a combined quantity such that
$\Delta m^{(1,2)}=|\Delta m^{(1)}-\Delta m^{(2)}|/\Delta m^{(1)}$. As a result we obtain an accuracy better than $1\%$ for any $n \leq 11$ while the expansion of
Jacobi polynomials contains nine terms to yield us a good approximation. This accuracy is obtained for both underlying pQCD and analytic perturbation theory
based on the  $2\delta$anQCD model. The numerical results for this accuracy at the energy scale $Q^2\approx 0.17\,{\rm GeV}^2$ are collected in Table
\ref{tab:Table2}.
\begin{table}[htb]
\caption{\footnotesize The accuracy in percent for the difference of the approximations
$\Delta m^{(1,2)}=\frac{|\Delta m^{(1)}-\Delta m^{(2)}|}{\Delta m^{(1)}}$
in the underlying pQCD: $\Delta m_{{\rm pQCD}}^{(1,2)}$; and for
anQCD in $2\delta$anQCD, $\Delta m_{{\rm anQCD}}^{(1,2)}$. \textcolor{black}{The results are presented for $Q^2\approx 0.17~{\rm GeV}^2$.}}
\label{tab:Table2}
\begin{center}
\centering
\begin{tabular}{llllll}
\hline
$n$ & 2 & 4 & 6 & 8 & 10
\\
\hline
$\Delta m_{{\rm pQCD}}^{(1,2)} \%$ & {1.372}  & ${0.817}$  & ${0.177}$  & ${0.368}$  & ${0.824}$  \\
\hline
$\Delta m_{{\rm anQCD}}^{(1,2)} \%$ & {0.499}  & ${0.296}$  & ${0.0642}$  & ${0.132}$  & ${0.296}$
\\
\hline
\end{tabular}
\end{center}
\end{table}
\begin{figure}[!htb]
	\vspace*{0.5cm}
	\includegraphics[clip,width=0.45\textwidth]{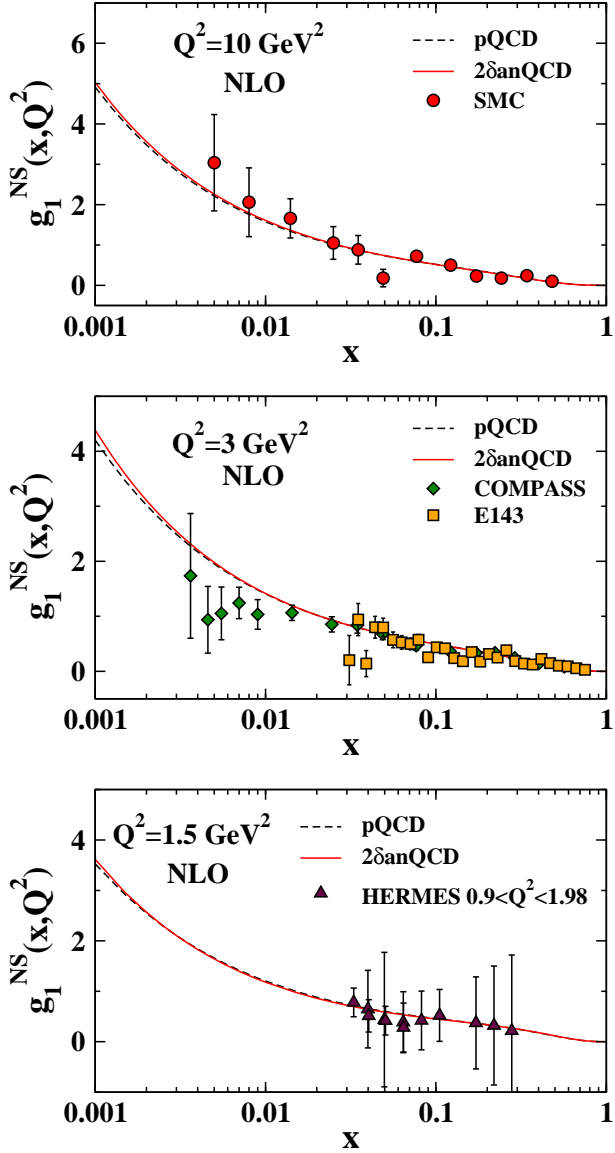}
	\begin{center}
		\caption{\small (Color online) Nonsinglet structure function $g_1^{(NS)}(x,Q^2)$ as a function of $x$ in NLO approximation. The solid line represents
the $2{\delta}$anQCD result, and the dashed line {{the underlying pQCD one}}. The available experimental data~\cite{Adeva:1998vv,Adolph:2015saz,Abe:1998wq,Ackerstaff:1997ws}
are included.  \label{fig:xgnsxdependent1}}
	\end{center}
\end{figure}

\begin{figure}[htb]
	\vspace*{0.5cm}
1	\includegraphics[clip,width=0.45\textwidth]{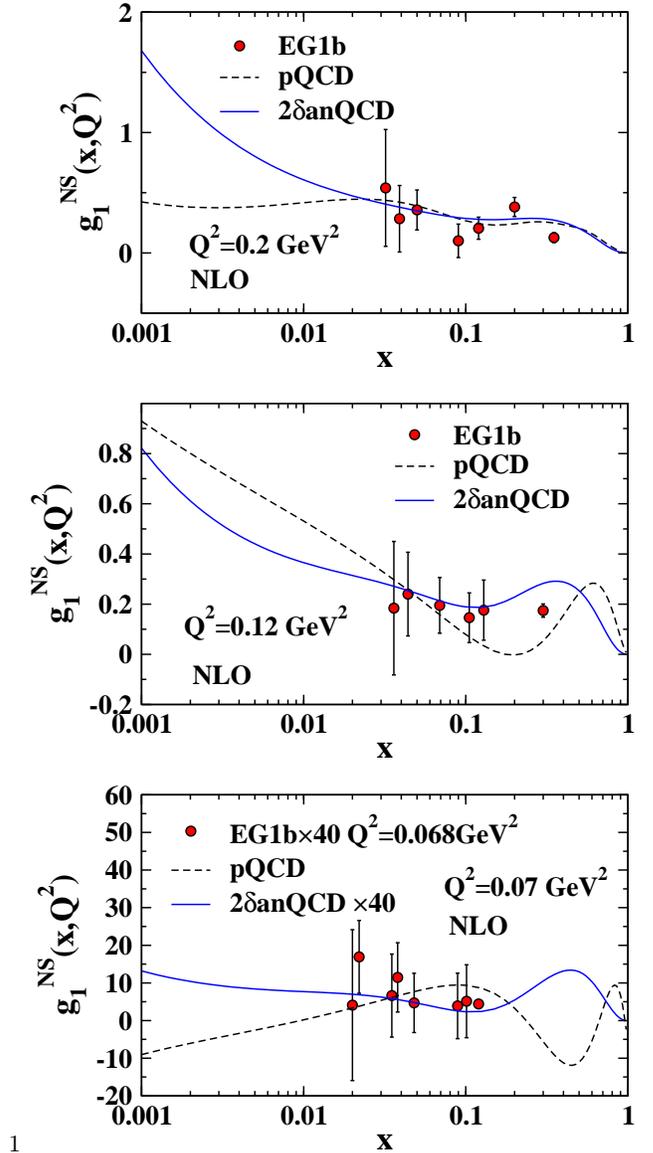}
	\begin{center}
		\caption{\small (Color online) Nonsinglet structure function $g_1^{(NS)}(x,Q^2)$ as a function of $x$ in the NLO approximation at low-energy scales. The
solid line with smooth behavior represents the $2{\delta}$anQCD result, and the dashed line {{ the underlying pQCD one}}. The available experimental data~\cite{priv}
are included.  \label{fig:xgnsxdependent}}
	\end{center}
\end{figure}
This confirms that the first rational approximation, $\Delta {\rm m}^{(1)}_\text{anQCD}(n,Q^2)$ in Eq.~(\ref{eq:mfapt}), has sufficient precision, i.e., the
corresponding Eq.~(\ref{eq:mnsapt}) gives us the desired results for the moments of the polarized nucleon structure function in the $2\delta$anQCD approach. The results, using these two approximations, are depicted in Fig.~\ref{fig:xgnsQdependent}, \textcolor{black} {i.e., the  polarized structure function $ g_1^{(NS)}$ as a function of $Q^2$ at $x=0.03$.} The plots in this figure denoted by $2\delta$anQCD$^{(1)}$ and $2\delta$anQCD$^{(2)}$,  which are related to the two
approximate expansions in Eqs.(\ref{eq:mfapt}) and ({\ref{eq:mfapt1}}), are in good agreement with each other and also  with the available experimental data. In this figure, some of data are from \cite{Airapetian:2006vy,Adeva:1998vv,Adolph:2015saz} \textcolor{black}{which have been collected in Table 1 of Ref.~\cite{Deur20019}}. The data at very low-energy scales, i.e. {{}$Q^2<0.5\; GeV^2$}, is from Ref.~\cite{priv}.
{\color{black} The plot for $g_1^{(NS)}$ at $x=0.03$ provides an opportunity to compare the available experimental data  at low-energy scales with the
theoretical prediction of 2$\delta$anQCD model, where the underlying pQCD does not give an acceptable behavior. This confirms the advantage and applicability  of the anQCD approach in comparison to pQCD, especially at the low-energy scales.}
In Fig.~\ref{fig:xgnsxdependent1} we present the results for  $g_1^{(NS)}$ as a function of $x$, at $Q^2= 10, 3, 1.5$ and $0.5 \ {\rm GeV}^2$ in 2$\delta$anQCD and
in the underlying pQCD, and  compare them with the available experimental data from SLAC \cite{Abe:1998wq}, HERMES \cite{Ackerstaff:1997ws}, SMC
\cite{Adeva:1998vv}  and COMPASS \cite{Adolph:2015saz} experimental groups. All these data have been collected in Table 1 of Ref.~\cite{Deur20019}. As expected, at high- and medium-energy scales, there is good  agrement between the
theoretical predictions and the experimental data.  Figure~\ref{fig:xgnsxdependent} is like  Fig.~\ref{fig:xgnsxdependent1} but at low-energy scales $Q^2=0.2,
0.12$ and $0.07 \ {\rm GeV}^2$. By decreasing the energy scales, the difference between the 2$\delta$anQCD and the underlying pQCD results increases. At $Q^2=0.07\; {\rm
GeV}^2$, the underlying pQCD result for $x\geq 0.1$ as a function of increasing $x$ grows rapidly and also oscillates. As a consequence, we multiplied there the
2$\delta$anQCD results \textcolor{black}{and the experimental results} by a factor of $40$ in order to facilitate the comparison with the pQCD results. Figure \ref{fig:xgnsxdependent} indicates that the
$g_1^{(NS)}$ structure function involves smooth behavior at very low-energy scales, using the 2$\delta$anQCD model. This can be considered as an advantage of this model in comparison with underlying pQCD.

{{}There are not enough experimental data  for $g_1^{(NS)}$ at the very low-energy scales $Q^2$. But there are individual experimental data for the polarized proton  and neutron structure functions, $g_1^p$ and $g_1^n$, respectively. Unfortunately, most of them are binned very differently with respect to the $x$-Bjorken variable and $Q^2$ scales and cannot be used  to construct the data points even utilizing the linear interpolation.  Nonetheless, there are limited data with the identical bins of $x$  and $Q^2$ and few others which are binned not so much differently with respect to the $x$ and $Q^2$  such that the linear interpolation method is applicable for them. These data are selected such as to be placed in the inelastic region. {{}In fact, these data are selected such that they exclude the elastic and $\Delta$ resonance particles}. They are depicted in three different panels in Fig.~\ref{fig:xgnsxdependent}.\footnote{We thank S.~E~K\"uhn for useful comments on this point.}}

\section{$Q^2$ dependence of Bjorken sum rule in anQCD approach}\label{sec:sumrull}

Here we investigate the Bjorken sum rule (BSR)~\cite{Bjorken:1966jh} in the $2\delta$anQCD model. This sum rule is relating the spin dependence of quark
densities to the axial charge.The BSR is important to understand  the nucleon spin structure,
that QCD can describe well the strong {{}force in the polarized case; i.e.} its $Q^2$ dependence reflects the strong force in the polarized case.
BSR has been measured at SLAC, DESY and CERN \cite{Abe:1997cx,Anthony:1999py,Anthony:1999rm,Anthony:2000fn,Abe:1997dp,Anthony:2002hy,Abe:1998wq,Adeva:1998vv,Ackerstaff:1997ws,Ackerstaff:1998ja,
Airapetian:1998wi,Airapetian:2002rw,Airapetian:2002wd,Airapetian:2006vy,Bosted:2001wm}
via polarized deep inelastic lepton scattering process. It has also been measured at
moderate values of $Q^2$  by Jefferson Lab (JLab)~\cite{Deur:2004ti,Deur:2008ej,Chen:2005tda,Deur:2014vea}. The $Q^2$ momentum that is probing the nucleon is related to the
inverse of the space-time scale. BSR  at high $Q^2$  was evaluated in conventional pQCD by Refs.~\cite{Kataev:1994gd,Kataev2005,Baikov2010,Pasechnik:2009yc} and has the form
\begin{eqnarray}
  &&\Gamma_1^{p-n} \equiv \int_{0}^{1} dx (g_1^p(x,Q^2)-g_1^n(x,Q^2))=\nonumber \\
   && \frac{g_A}{6}\left[1-\frac{\alpha_s}{\pi}-d_1\frac{\alpha_s^2}{\pi^2}-d_2\frac{\alpha_s^3}{\pi^3}+d_3\frac{\alpha_s^4}{\pi^4}+\ldots\right]+\nonumber\\
   &&\sum_{i=2}^{\infty}\frac{\mu_{2i}^{p-n}(Q^2)}{Q^{2i-2}}\,.\nonumber\\ \label{eq:sumrull}
\end{eqnarray}
In this equation the spin-dependent proton and neutron structure functions are $g^p_1$ and $g^n_1$, respectively. The strength of the neutron $\beta$ decay is
controlling by $g_A$ which is the nucleon axial charge.  Soft gluon radiation in conventional pQCD causes  {{} the leading-twist term (known as $\mu_2$), given by the first term in the right-hand side of Eq.~(\ref{eq:sumrull})}, to depend smoothly on $Q^2$. The $d_i$ coefficients up to $\alpha_s^4$ order can be found in Ref.~\cite{Baikov2010}. The terms with $\mu_4$, $\mu_6$, etc.,
are the nonperturbative power corrections, usually known as higher twists (HT). The higher twists, in fact, reflect a correlation between quarks and gluons. For a good understanding of the nucleon structure at the low-energy scales, analyzing  the HT is essential. This is why it is interesting to follow as well this
subject in the anQCD approaches, as we mentioned in the introduction. We recall that in the introductory part, a brief review  has been done independent of the specific scheme. Therefore, the references cited here are not also in the Lambert scheme \textcolor{black}{(usually they are in the ${\overline {\rm MS}}$ scheme), but the evaluations here are in the ($c_2=-4.9$) Lambert scheme.}

The anQCD modification of the BSR , using the 2$\delta$anQCD model,  has the form
\begin{eqnarray}\label{eq:sumrulfapt}
  \Gamma_1^{p-n} = \Gamma_{1,anQCD}^{p-n}+\sum_{i=2}^{\infty}\frac{\mu^{anQCD}_{2i}(Q^2)}{Q^{2i-2}}\,
\end{eqnarray}
where
\begin{eqnarray}
\Gamma_{1,anQCD}^{p-n} = \frac{g_A}{6}\left[1-\Delta^{p-n}_{1,anQCD}(Q^2)\right]\;,\nonumber
\end{eqnarray}
and
\begin{eqnarray}
\Delta^{p-n}_{1,anQCD} =d_1(k){\cal A}_1+d_2(k){\cal A}_2
  +d_3(k){\cal A}_3+d_4(k){\cal A}_4,\nonumber
\end{eqnarray}
{\color{black}where $d_i{(k)}$ coefficients are in the Lambert scheme with $c_2=-4.9$ (cf.~Appendix A in Ref.\cite{cv-ay-2018}). The $k$ parameter is the renormalization scale parameter $k = \mu^2/Q^2$, the couplings are ${\cal A}_n ={\cal A}_n^{(2 \delta)}(\mu^2)$, and we fixed the value of $k$ to $k=1$.}

The higher-twist effects are included in Eq.~(\ref{eq:sumrulfapt}). The term  with dimension $D = 2$, i.e., $\mu_4^{p-n}/Q^2$, has the following coefficient
\cite{sh}:
\begin{equation}\label{eq:mu}
\mu_4=\frac{M_N^2}{9}(a_2^{p-n}+4d_2^{p-n}+4f_2^{p-n}(Q^2)) .
\end{equation}
In this equation the nucleon mass is $M_N \approx 0.94\; {\rm GeV}$. The coefficient  $a_2^{p-n}$ represents the twist-2 target mass correction; $d_2^{p-n}$ is
related to the twist-3 matrix element \cite{sh}. These coefficients can be computed, using the following relations \cite{sh}:
\begin{eqnarray}\label{eq:a2d2}
  a_2^{p-n} &=&\int_{0}^{1}dxx^2g_1^{p-n}\;,\nonumber \\
  d_2^{p-n} &=&\int_{0}^{1}dxx^2(2g_1^{p-n}+3g_2^{p-n})\;.
\end{eqnarray}

{{} To calculate $d_2^{p-n}$ we need the $g_2$ structure function. However, there is a relation which gives us $g_2$ in terms of the $g_1$ structure function in Ref.~\cite{FF52} but this relation includes just the twist-2 part of $g_2$ and it cannot be substituted in Eq.(\ref{eq:a2d2}) since it is related to the twist-3 matrix element. Instead, one can resort to the quoted value for $d_2^{p-n}$ in Refs.~\cite{Deur:2014vea,Deur20019}, a result from recent experimental analysis. The numerical value which is reported there, is $d_2^{p-n}=0.008\pm 0.0036$.

Since one can compute the polarized structure functions $g_1^{p}$ and $g_1^{n}$ [see Eq.~(\ref{eq:eg1Jacob})], it is possible to obtain the numerical values for $a_2^{p-n}$  in Eq.~(\ref{eq:a2d2}) at $Q^2=1\; {\rm GeV}^2$. i.e. $a_2^{p-n}=0.0157\pm 0.0001$. To calculate the polarized structure function $g_1^{NS}$  it is required at first to achieve the polarized valence densities at initial scale $Q_0$ , given by Eq.(\ref{eq:parametrizationsQ0}). The unknown parameters of these densities can be obtained using the traditional global fit which has been done in Refs.~\cite{Khorramian:2010qa,Khanpour:2017fey}. It should be recalled that  the polarized structure function $g_1^{NS}$  can be calculated, based on the formalism which we applied in this article. {{}But the numerical values of  the unknown parameters for the required parton densities have been quoted from Refs.~\cite{Khorramian:2010qa,Khanpour:2017fey}, listed in Table \ref{tab:fit}.}

The $f_2^{p-n}$ function in Eq.(\ref{eq:mu}) is the \textcolor{black}{expectation value of} a well-defined operator with a specific physical meaning \cite{bu,ab}. Here it is regarded as a parameter which parameterizes a power correction to the anQCD analysis. From this point of view, we take the anQCD evolution form (based on its pQCD form) such that $f_2^{p-n}(Q)=f_2^{p-n}(Q_0) {\A}^{(2\delta)}_{\frac{\gamma_0}{8\beta_0}}(Q^2)/{\A}^{(2\delta)}_{\frac{\gamma_0}{8\beta_0}}(Q^2_{0})$ as has been quoted in Ref.~\cite{cv-ay-2018}.

While $f_2^{p-n}$ is evaluated in the 2$\delta$anQCD approach, the coefficients $a_2$ and $d_2$, as twist-2 and twist-3 quantities,  are obtained from the underlying pQCD   and the analysis of related experimental data, respectively. In fact the first integral in Eq.~(\ref{eq:a2d2}) {can be calculated from the computed polarized structure function $g_1^{NS}$. As we explained in the two previous  paragraph,
to compute it one needs to use the traditional global fit as in Refs.~\cite{Khorramian:2010qa,Khanpour:2017fey}} to obtain the unknown parameters of the polarized valence densities in Eq.(\ref{eq:parametrizationsQ0}).

We extend our computations up to the fourth higher twist to include as well the  $\mu_6$ term. In our calculations we
have in total two free parameters  $\mu_6$ and $f_2^{p-n}(Q_0)$ which can be obtained by fitting to the available experimental data for the BSR. The fitted numerical values are  $\mu_6=0.0007\pm0.0001$ and $f_2^{p-n}(Q_0)=-0.020\pm 0.001$ ,respectively. Based on Eq.(\ref{eq:mu}) the numerical value for $\mu_4$ would be $\mu_4=-0.0031\pm0.001$.}

\textcolor{black}{Equation (\ref{eq:sumrulfapt}) is the analytical result for $\Gamma_1^{p-n}$ in the $2\delta$anQCD model.}
\begin{figure}[!htb]
	\vspace*{0.5cm}
	\includegraphics[clip,width=0.4\textwidth]{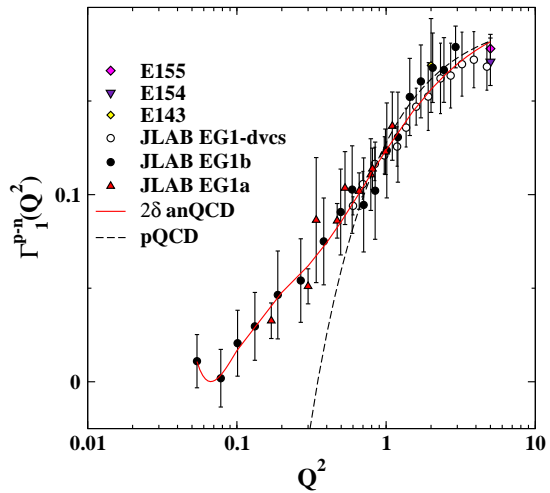}
	\begin{center}
		\caption{\small (Color online)The BSR, $\Gamma_1^{p-n}(Q^2)$, resulted from two models.The
 red line gives the result from 2$\delta$anQCD model and black line from  the underlying pQCD. Other symbols show the data from  E143 , E154 and
 E155~\cite{Abe:1998wq,Abe:1997dp,Bosted:2001wm} and JLab~\cite{Deur:2004ti,Deur:2008ej,Chen:2005tda,Deur:2014vea} experiments.\label{fig:firstmoment}}
	\end{center}
\end{figure}
In Fig.~\ref{fig:firstmoment}, we present $\Gamma_1^{p-n}$ in the $2\delta$anQCD approach and compare with the result from {{{} {the underlying pQCD where  both are considered in the Lambert scheme}}, as well as  with the E143 , E154
and E155~\cite{Abe:1998wq,Abe:1997dp,Bosted:2001wm} and JLab~\cite{Deur:2004ti,Deur:2008ej,Chen:2005tda,Deur:2014vea} experimental data.
The good agreement between the available experimental data and the analytical result from the 2$\delta$anQCD model, in contrast to the underlying pQCD result
especially  at low energies,  confirms that the anQCD approach, based on the employed model, is working well.
\section{SUMMARY and Conclusion}\label{summery}
As a new  theoretical approach to analyze {{} the nonsinglet polarized structure function at the low $Q^2$ continuum region, the
analytic QCD (anQCD), specifically the 2$\delta$anQCD model, is employed to evaluate this function and also  the $Q^2$ dependence of Bjorken sum rule.}

Using this approach, it is possible to analyze the polarized  structure function in the whole $Q^2$ range at the leading-twist order. On this basis, one may
conclude some outstanding  characteristics arising from the anQCD approach. At first, it can be shown that the $g_1^{(NS)}(x,Q^2)$ structure function at a fixed
$x$ value and in the whole range of $Q^2$ is slowly changing. The plot in Fig.~\ref{fig:xgnsQdependent} reveals this feature. Second, the results of the underlying pQCD
and anQCD approaches at moderate and high $Q^2$ scales are consistent with each other as can be seen from the depicted plots in Fig.~\ref{fig:xgnsxdependent1}.
Third, at low values of $Q^2$, the evolution of $g_1^{(NS)}(x,Q^2)$ as a function of $x$ in the  anQCD approach is slower and smoother in comparison with the
{{{} underlying pQCD which are both in the Lambert scheme}}. This fact can be seen from the three plots in Fig.~\ref{fig:xgnsxdependent} at the low-energy scales $Q^2$=0.2, 0.17 and $0.07 \ {\rm
GeV}^2$, respectively. {{} Consequently, we conclude that the result from  analytic series in the anQCD approach, using the 2$\delta$anQCD model,  better reproduces the data
and behaves more smoothly  than in underlying pQCD, especially at low $Q^2$  values.}

We also investigated the $Q^2$ dependence of the Bjorken sum rule in the anQCD approach, using the 2$\delta$anQCD model. The result is shown in
Fig.~\ref{fig:firstmoment} which indicates good agreement with the available experimental data in the whole interval of energy scales. The agreement between them
especially at the low-energy scales confirms the validity of the employed anQCD approach. A comparison has also been done with the result from the underlying
pQCD approach, suggesting an advantage of the {{}anQCD  over the pQCD approach}} in the evaluation of spacelike QCD observables at low energies .

All these features have their origin in the fact that the behavior of analytic (holomorphic) coupling is under control in the entire range of spacelike $Q^2$
values, including at $Q^2 \mapsto 0$. This outstanding feature has been presented in Fig.~\ref{fig:A-1}.

As a final point we mention that the plots in Fig.~\ref{fig:xgnsQdependent}, {{} toward small values of $Q^2$}},  resulted from  Eq.~(\ref{eq:mnsapt}) in the 2$\delta$anQCD model. In this case the
Jacobi polynomial expansion is used [see Eq.~(\ref{eq:eg1Jacob})] to convert the result from the moment space to $x$-Bjorken space. In Eq.~(\ref{eq:eg1Jacob})
,the notation ${\Delta M}[xg_{1}^{NS}, j+2]$  for the polarized moment of structure function is equivalent to the $\Delta {\cal M}_\text{NS}(n,Q^2)$ in
Eq.~(\ref{eq:mnsapt}), \textcolor{black}{with $n=j+2$.} It should be noted that in converting the structure function from the moment space to Bjorken-$x$ space in underlying pQCD,  we used the inverse Mellin technique based on Eq.~(\ref{eq:inversmellin}),  while for the anQCD approach, due to computational difficulties
which arise in employing inverse Mellin technique, we resorted to employing the technique of Jacobi polynomial expansion.\\\\
A further research task and an interesting subject would be to extend the anQCD analysis to the nonsinglet case for the unpolarized structure function which
contains heavy quark flavors. {{} In this case  the effect of quark mass cannot be ignored, and it should  be considered in the calculations.  In this regard the related commands in the 2$\delta$anQCD model would be different with respect to what we used here for light quarks.}
Also of interest would be to consider the singlet case of structure functions in the anQCD approach, a subject on which we hope to
report in the future.

\section*{ACKNOWLEDGMENTS}
The authors are indebted  to G.~Cveti\v{c} and C.~Ayala  for reading the manuscript and  providing crucial comments which helped us improve the manuscript.  We
are grateful to S.~E.~K\"uhn for the very essential discussions  about the CLAS data. We also acknowledge  G.~Mallot for productive comments and providing us with the required COMPASS data. We are finally thankful to Charles C. Young for his useful comments in connection to the related CERN data. S.A.T is grateful to the School of Particles and Accelerators, Institute for Research in Fundamental Sciences (IPM) to make the required facilities to do this project.

%
%
\section*{Appendix A: The splitting functions in the NLO approximation}\label{Sec:AppendixA}
The analytical expression for  the splitting functions in the moment space for nonsinglet sectors and at NLO approximation are taken from  \cite{Lampe:1998eu}
and presented below. The usual quadratic Casimir operators are fixed to their exact values, using $C_A = 3$, ${\rm T_F} = \frac{1}{2}$ and $C_F = \frac{4}{3}$.
\begin{widetext}
 	\begin{eqnarray}
-\Delta P_{NS \pm}^{(1)n}&=&
   C_F^2 \Bigg[ 2\frac{2n+1}{n^2 (n+1)^2}S_1(n)
         +2(2S_1(n)-\frac{1}{n(n+1)})(S_2(n)-S_2'(\frac{n}{2}))
\nonumber \\
& &          +3S_2(n)+8\tilde{S}(n)-S_3'(\frac{n}{2})
         -\frac{3n^3+n^2-1}{n^3 (n+1)^3}-\frac{3}{8}
         \mp 2\frac{2n^2+2n+1}{n^3 (n+1)^3} \Bigg]
\nonumber \\
& &   +C_F C_A \Bigg[\frac{67}{9}S_1(n)
             -(2S_1(n)-\frac{1}{n (n+1)})(2S_2(n)-S_2'(\frac{n}{2}))
             -\frac{11}{3}S_2(n)-4\tilde{S}(n)+\frac{1}{2}S_3'(\frac{n}{2})
\nonumber \\
& &        -\frac{1}{18}\frac{151n^4+236n^3+88n^2+3n+18}{n^3 (n+1)^3}
       -\frac{17}{24}\pm \frac{2n^2+2n+1}{n^3 (n+1)^3} \Bigg]
\nonumber \\
& & +C_F T_f\Bigg[-\frac{20}{9}S_1(n)+\frac{4}{3}S_2(n)
           +\frac{2}{9}\frac{11n^2+5n-3}{n^2 (n+1)^2}+\frac{1}{6}\Bigg]\;,\nonumber\hspace{6 cm}\text{(A.1)}
\/ .
\label{eq:a010}
\end{eqnarray}
where
\begin{eqnarray}\label{eq:polygamma}
S_k(n) & \equiv & \sum_{j=1}^n \frac{1}{j^k}\;,\nonumber\hspace{14 cm}\text{(A.2)}\\
S_k'\left(\frac{n}{2}\right) & \equiv & 2^{k-1} \sum_{j=1}^n
\frac{1+(-)^j}{j^k}
= \frac{1}{2} (1+\eta ) S_k\left(\frac{n}{2}\right)+
\frac{1}{2} (1-\eta ) S_k\left(\frac{n-1}{2}\right)\;,\nonumber\hspace{5.75 cm}\text{(A.3)}\\
\tilde{S}(n) & \equiv & \sum_{j=1}^n \frac{(-)^j}{j^2} S_1(j)
=  -\frac{5}{8} \zeta (3) +\eta \left[ \frac{S_1(n)}{n^2} +
\frac{\pi^2 }{12} G(n) +\int_0^1 dx\; x^{n-1} \frac{{\rm{Li}}_2(x)}
{1+x}\right]\;,\nonumber\hspace{4.35 cm}\text{(A.4)}\\
\frac{{\rm{Li}}_2(x)}{(1+x)}& \equiv &(1.01 x - 0.846 x^2 + 1.155 x^3 - 1.074 x^4 + 0.55 x^5)\;.\nonumber\hspace{8 cm}\text{(A.5)}
\end{eqnarray}
\end{widetext}
In Eq.~(A.4), $G(n)$ is defined as $G(n)\equiv \psi\left(\frac{n+1}{2}\right) -
\psi\left(\frac{n}{2}\right)$ where $\psi (z)=d\ln \Gamma (z) /dz$. For $\Delta P_{NS \pm}^{(1)n}$  it is assumed $\eta =\pm 1$ while
for  anomalous dimensions of  nonsinglet flavor,  $\eta =+ 1$ is considered.

The $S_k(n)$ functions can be written in terms of the harmonic sums  such as~\cite{Vermaseren:1998uu,Blumlein:1998if},
\begin{eqnarray}
S_1(n) &=& \gamma _E + \psi (n+1)\,, \nonumber\\
S_2(n) &=& \zeta (2) - \psi  '(n+1)\,, \nonumber\\
S_3(n) &=& \zeta (3) + 0.5 \, \psi''(n+1)\,, \nonumber\hspace{2 cm}\text{(A.6)}
\end{eqnarray}

where $\gamma _E = 0.577216$ is the Euler constant , $\zeta (2) = \pi^2/6$ and $\zeta (3) = 1.20206$ .

\end{document}